\documentclass{article}
\usepackage[dvips]{epsfig}

\begin{document}
\title{Cryptanalysis of the RSA-CEGD protocol}
%\title{Unfairnes of a protocol for certified delivery}
\author{Juan M. E. Tapiador, Almudena Alcaide,\\
Julio C. Hernandez-Castro, and Arturo Ribagorda\\[2ex]
Computer Science Department, Carlos III University of Madrid\\
Avda. Universidad 30, 28911, Legan\'es, Madrid (Spain)\\
\texttt{\{jestevez, aalcaide, jcesar, arturo\}@inf.uc3m.es}}

\date{}
\maketitle
%
%
%------------------------------------------------------------------------------
\begin{abstract}
Recently, Nenadi\'c \emph{et al.} (2004) proposed the RSA-CEGD
protocol for certified delivery of e-goods. This is a relatively
complex scheme based on verifiable and recoverable encrypted
signatures (VRES) to guarantee properties such as strong fairness
and non-repudiation, among others. In this paper, we demonstrate how
this protocol cannot achieve fairness by presenting a severe attack
and also pointing out some other weaknesses.\\

\end{abstract}
%------------------------------------------------------------------------------
%
%
%
%------------------------------------------------------------------------------
\section{Introduction}\label{Sec:Introduction}
%------------------------------------------------------------------------------
%
Interest in protocols for fair exchange of information with
non-repudiation stems from its importance in many applications where
disputes among parties can occur. Assurance of these properties
enables the deployment of a wide range of applications, such as
certified e-mail or business transactions through communication
networks. As a result, fair non-repudiation has experienced an
explosion of proposals in recent years (see \cite{KMZ02} for an
excellent survey).

Nevertheless, fairness and non-repudiation have not been so
extensively studied as other classic issues, such as confidentiality
or authentication. Previous experience in these contexts has shown
that designing security protocols is an error-prone task. Consider,
as an illustrative example, a non-repudiation protocol proposed in
1996 by Zhou and Gollman \cite{ZG96} that was verified and proved
correct using three different methods \cite{BP01,Sch98,ZG98}.
Surprisingly, in 2002 G\"urgens and Rudolph demonstrated the absence
of fair non-repudiation in that protocol under reasonable
assumptions \cite{GR02}. In this case, possible attacks were
detected after an analysis performed with a different formalism that
considered scenarios not checked before.

The RSA-CEGD protocol \cite{NZCG04} was recently proposed for
certified delivery of e-goods, i.e. commercial products that can be
represented in electronic form and transmitted over open networks.
The scheme is designed to satisfy six major security requirements:
non-repudiation of origin; non-repudiation of receipt; strong
fairness; e-goods content/quality assurance; e-goods and receipt
confidentiality; and transparency of the STTP. In this paper we
demonstrate that this protocol suffers from severe security
problems, and some of the requirements mentioned above cannot be
satisfied. In particular, we present attacks that show how the
protocol does not assure fairness.

The rest of this paper is organized as follows. Section
\ref{Sec:RSA-CEGD} introduces the notation and briefly reviews the
RSA-CEGD protocol. Section \ref{Sec:Vulnerabilities} discusses the
vulnerabilities and illustrate them through specific attack
scenarios. Finally, Section \ref{Sec:Conclusions} summarizes the
paper by presenting some conclusions.

%
%==============================================================================
\section{Overview of the RSA-CEGD protocol}\label{Sec:RSA-CEGD}
%==============================================================================
%
For readability and completeness, we first provide a brief review of
the RSA-CEGD protocol.

%
%.........................................................................
\subsection{Notation}
%.........................................................................
%
Throughout this paper, we will use the same notation introduced by
the authors in the original paper \cite{NZCG04}. The protocol's
items and cryptographic symbols are described below.

\begin{itemize}
    \item $P_a$, $P_b$, $P_t$: different protocol parties, where
    $P_a$ is the e-goods provider (\emph{message sender}) and $P_b$ is the
    purchaser (\emph{message receiver}). $P_t$ acts as a Semi-Trusted
    Third Party (STTP).
    \item $D_a$: e-goods to be purchased.
    \item $k_a$: symmetric key used by $P_a$ to encrypt $D_a$.
    \item $r_a$ : random prime generated by $P_a$.
    \item $x_a = (r_a \times k_a)\mbox{ }mod\mbox{ }n_a$ : encryption of
    key $k_a$ with random number $r_a$
    \item $CertD_a = (desc_a, hd_a, h_a, ek_a, sign_{CA})$ : certificate for
    $D_a$ issued by a CA, where:
    \begin{itemize}
       \item $desc_a$ = description (content summary) of $D_a$
       \item $hd_a = h(E_{k_a}(D_a))$ : hash value of the encryption of
       $D_a$ with key $k_a$
       \item $h_a = h(D_a)$ : hash value of $D_a$
       \item $ek_a = E_{pk_a}(k_a)$ : encryption of the key $k_a$
       with $P_a$'s public key, $pk_a$
    \end{itemize}
    \item $E_{sk_a}(h_a)$ : $P_a$'s RSA signature on $D_a$ serving as a proof
    of origin of $D_a$
    \item $y_a = E_{pk_a}(r_a)$ : RSA encryption of number $r_a$ with key $pk_a$
    \item $r_b$ : random prime generated by $P_b$ for the generation of the
    VRES $(y_b, x_b, xx_b)$.
    \item $rec_b = (h_a)^{d_b} \mbox{ } mod \mbox{ } n_b$ : $P_b$'s receipt
    for $P_a$'s e-goods $D_a$, i.e. $P_b$'s RSA signature on $D_a$
    \item $(y_b, x_b, xx_b)$ : $P_b$'s VRES, where
    \begin{itemize}
       \item $y_b = {r_b}^{e_b} \mbox{ } mod \mbox{ } (n_b \times n_{bt})$ :
       encryption of $r_b$ with $P_b$'s public key. Also recoverable by $P_t$
       \item $x_b = (r_b \times (h_a)^{d_b}) \mbox{ } mod \mbox{ } n_b = (r_b
       \times rec_b) \mbox{ } mod \mbox{ } n_b$ : encryption of $rec_b$ with $r_b$
       \item $xx_b = (r_b \times E_{sk_{bt}}(h(y_b)))\mbox{ } mod \mbox{ }
       n_{bt}$ : control number that confirms the correct use of $r_b$
    \end{itemize}
    \item $C_{bt} = (pk_{bt}, w_{bt}, s_{bt})$ : $P_b$'s RSA public-key certificate issued by $P_t$
    \item $pk_{bt} = (e_{bt}, n_{bt})$ : public RSA key related to $C_{bt}$, with $e_{bt} = e_b$
    \item $sk_{bt} = (d_{bt}, n_{bt})$ : private RSA key related to $C_{bt}$
    \item $w_{bt} = ( h(sk_t, pk_{bt})^{-1} \times d_{bt})\mbox{ }mod\mbox{ }n_{bt}$
    \item $s_{bt} = E_{sk_{t}}(h(pk_{bt}, w_{bt}))$ : $P_t$'s signature on $h(pk_{bt}, w_{bt})$
    \item $s_b = E_{sk_b}(h(C_{bt}, y_b, y_a, P_a))$ : $P_b$'s recovery authorization token.
\end{itemize}

%
%.........................................................................
\subsection{Exchange and recovery sub-protocols}
%.........................................................................
%
The RSA-CEGD is an optimistic fair exchange protocol composed of two
sub-protocols, as shown in Fig. \ref{Fig:RSA-CEGD}. As usual, the
exchange sub-protocol is used to carry out the exchange between
parties without any TTP's involvement. In case the process fails to
complete successfully, a recovery protocol can be invoked to handle
this situation.

%:::::::::::::::::::: Figure: RSA-CEGD Protocol :::::::::::::::::::::::::::::
\begin{figure}[t]
\begin{center}
\begin{tabular}[3]{lll}

\hline

\hline

\multicolumn{3}{l}{\textbf{The exchange sub-protocol}}\\

\hline

E1: & $\mathtt{P_a} \rightarrow \mathtt{P_b}:$ & $E_{k_a}(D_a),
\mbox{ } CertD_a, \mbox{ } x_a, \mbox{ } E_{sk_a}(h_a)$ \\

E2: & $\mathtt{P_b} \rightarrow \mathtt{P_a}:$ & $(x_b, xx_b,
y_b),
\mbox{ } s_b, \mbox{ } C_{bt}$ \\

E3: & $\mathtt{P_a} \rightarrow \mathtt{P_b}:$ & $r_a$ \\

E4: & $\mathtt{P_b} \rightarrow \mathtt{P_a}:$ & $r_b$ \\

\hline

\hline

\\

\hline

\hline

\multicolumn{3}{l}{\textbf{The recovery sub-protocol}}\\

\hline

R1: & $\mathtt{P_a} \rightarrow \mathtt{P_t}:$ & $C_{bt},
\mbox{ } y_b, \mbox{ } s_b, \mbox{ } y_a, \mbox{ } r_a$ \\

R2: & $\mathtt{P_t} \rightarrow \mathtt{P_a}:$ & $r_b$ \\

R3: & $\mathtt{P_t} \rightarrow \mathtt{P_b}:$ & $r_a$ \\

& & \\

\hline

\hline
\end{tabular}
\end{center}
\caption{The RSA-CEGD protocol.}\label{Fig:RSA-CEGD}
\end{figure}
%::::::::::::::::::::::::::::::::::::::::::::::::::::::::::::::::::::::::::::

The notion of verifiable and recoverable encrypted signature (VRES)
underlies at the core of the RSA-CEGD protocol. A VRES is basically
an encrypted signature, which acts as a \emph{receipt} from the
receiver's point of view, with two main properties. First, it can be
\emph{verified}: the receiver is assured that the VRES contains the
expected signature without obtaining any valuable information about
the signature itself during the verification process. And second,
the receiver is assured that the original signature can be
\emph{recovered} with the assistance of a designated TTP in case the
original sender refuses to do it.

Due to these two properties, the VRES becomes an interesting
cryptographic primitive upon which fairness can be provided. The
RSA-CEGD protocol relies on this element within the general scheme
we sketch in what follows:

\begin{enumerate}
   \item $A$ ciphers the message with an encryption key and sends it to $B$.
   \item $B$ generates the VRES of his signature and sends it back to $A$.
   \item Upon successful verification of the VRES, $A$ is
   assured that it is secure for her to send the decryption key to
   $B$, so he can access the message.
   \item Finally, $B$ sends his original signature to $A$ as a
   receipt. In case he refuses, a TTP can recover the
   signature from the VRES, thus restoring fairness.
\end{enumerate}

The RSA-CEGD protocol makes use of a novel VRES method based on the
RSA system, hence its name. The idea stems from the so-called
\emph{theory of cross-decryption} \cite{RR00}, which establishes
that an RSA encrypted text can be decrypted by using two different
keys if both pairs of secret/public keys are appropriately chosen.
Party $B$ is enforced to use a key of this kind to encrypt the VRES,
while the TTP retains the other. This way, if subsequently $B$
refuses to provide $A$ with his signature, the TTP is able to
recover it from the VRES.

%
%------------------------------------------------------------------------------
\section{Protocol vulnerabilities}\label{Sec:Vulnerabilities}
%------------------------------------------------------------------------------
%
Before stating specific attack scenarios, note that:
\begin{enumerate}
    \item The VRES received by party $P_a$ in step E2 contains the
    receipt $rec_b$, though it is not directly accessible to her.
    However, party $P_a$ is provided with \emph{all}
    the information required by the STTP to assist $P_a$ in the
    recovery of the receipt, i.e. the authorization token $s_b$ and
    $P_b$'s certificate $C_{bt}$.
    \item Items $<(x_b, xx_b, y_b), s_b, C_{bt}>$ do not contain
    themselves any link to the current protocol execution. They only
    refer to the e-goods $D_a$, the receipt $rec_b$, an authorization to
    $P_a$, $P_b$'s certificate, and the random numbers $r_a$ and $r_b$.
    \item The STTP can restore fairness only upon $P_a$ request.
    Party $P_b$ has no means to invoke a recovery sub-protocol.
    This puts $P_a$ in an advantageous position with respect to the
    other party.
    \item The protocol defines a token for non-repudiation of origin which
    does not include information to verify who submitted such a token to
    $P_b$.

\end{enumerate}

Invocation of the recovery sub-protocol by party $P_a$ will provide
$P_b$ with the number $r_a$, thus being able to recover the
encryption key and, hence, access the e-goods $D_a$. Nevertheless,
$P_a$ can appeal to the STTP during a \emph{different} protocol
execution, since the information required to access the receipt does
not identify the protocol session. In this scenario, the recovery
sub-protocol also sends number $r_a$ to $P_b$. However, the protocol
specification does not require $P_b$ to try the key received on
messages of previous exchanges. In other words, is not reasonable to
assume that $P_b$ stores all proofs he ever received, especially
those related to previous, unsuccessful exchanges.

As a result of the scheme outlined above, party $P_a$ obtains a
valid proof (receipt) of $P_b$ having received e-goods $D_a$. $P_b$,
on the other hand, does not have access to e-goods $D_a$ (or is not
aware that he has received the correct decryption key). Thus,
non-repudiation is not satisfied and the protocol does not provide
fairness for $P_b$. This situation is described in detail in the
attack scenario described in the following section. Furthermore,
some other weaknesses are pointed out in Section \ref{Sec:EOO}.

%
%.........................................................................
\subsection{A replay attack}\label{Sec:ReplayAttack}
\begin{figure}[t]
\begin{center}
\includegraphics[scale=0.4]{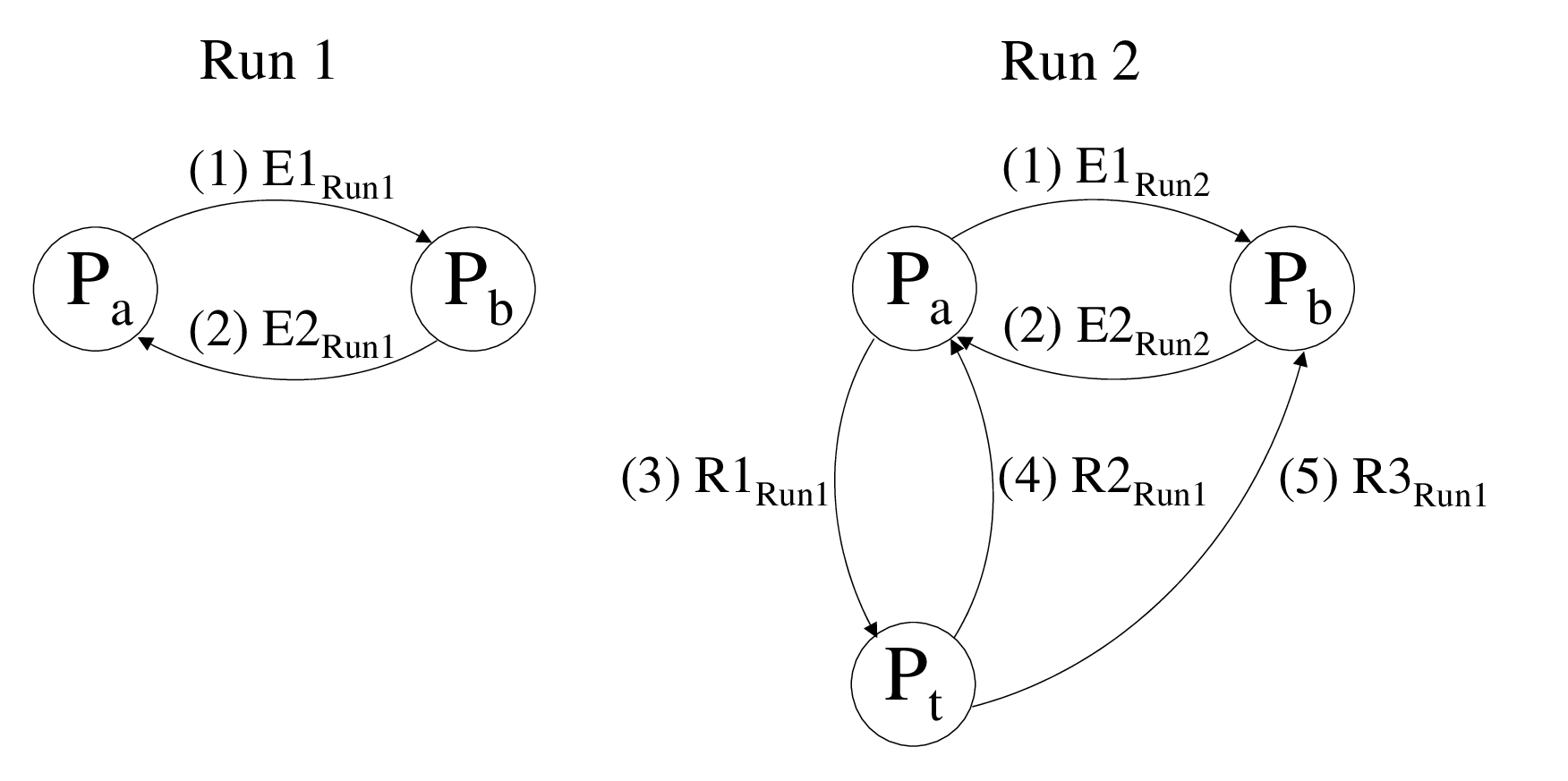}
\end{center}
\caption{Scheme of the attack.} \label{Fig:SchemeAttack}
\end{figure}

%.........................................................................
%
The basic scenario is graphically sketched in Fig.
\ref{Fig:SchemeAttack}. The attack is executed through two different
protocol runs between the same parties, $P_a$ and $P_b$. This is not
a strong assumption, since it is reasonable to expect that $P_b$
wishes to buy several e-goods from the same seller.

During the first protocol running, $P_a$ carries out step E1 and
then waits for the VRES, the authorization token, and $P_b$'s
certificate. We assume that $P_a$ performs the required
verifications on these items, so she is assured they are valid. At
this point, $P_a$ aborts the protocol. In fact, there is no abort
procedure per se, so she only does not continue with step E3. Note
as well that $P_b$ has no means to invoke a recovery sub-protocol in
this situation.

Now $P_a$ owns the received items:
\begin{displaymath}
<(x_b, xx_b, y_b), s_b, C_{bt}>
\end{displaymath}

\noindent and also number $r_a$ and its signature, $y_a$. From
these, $P_a$ constructs and stores the following message:
\begin{displaymath}
m_1 = <C_{bt}, y_b, s_b, y_a, r_a>
\end{displaymath}

Suppose that subsequently $P_b$ contacts $P_a$ to initiate another
exchange aimed at buying a different e-good, say $D_a'$. Again,
$P_a$ follows step E1 and, after E2, she receives:
\begin{displaymath}
<(x_b', xx_b', y_b'), s_b', C_{bt}>
\end{displaymath}

\noindent from $P_b$. Then, $P_a$ aborts the exchange sub-protocol
and starts an instance of the recovery sub-protocol. According to
the protocol semantics, it is expected that $P_a$ sends the
following items to the STTP in step R1:
\begin{displaymath}
m_2 = <C_{bt}, y_b', s_b', y_a', r_a'>
\end{displaymath}

\noindent However, $P_a$ chooses $m_1$ as the message to send. As
this is a valid proof, the STTP will recover numbers $r_b$ and
$r_a$, which will be sent to $P_a$ and $P_b$, respectively. The key
point is that both numbers are not related to the current protocol
execution, but with the previous one. This way, $P_a$ can use $r_b$
to obtain the receipt $rec_b$ contained in $m_1$. Even though $P_b$
also receives $r_a$, this number is useless for him to recover the
key required to access $D_a'$. In fact, this $r_a$ might provide
$P_b$ with access to the former e-goods he tried to buy. However, in
all likelihood he is not aware of this.

As a result, $P_a$ has a valid receipt of $P_b$ having received
e-goods $D_a$, though $P_b$ does not actually own it. Therefore, the
protocol does not provide fairness for $P_b$.

%
%.........................................................................
\subsection{Indistinguishability of evidences of origin}\label{Sec:EOO}
%.........................................................................
%
The protocol establishes the item $E_{sk_a}(h_a)$ as proof of
non-repudiation of origin. However, parties' identities are not
included in such a token, nor any other information related to the
current protocol execution. Even using authenticated channels,
evidences obtained do not link together the sender, the originator,
the receiver, the current protocol execution, etc. This fact yields
to a weakness related to the indistinguishability of evidences
exchanged during the protocol, in particular, evidence of origin
(EOO).

Suppose $P_a$ and $P_b$ perform a protocol execution, so finally
$P_b$ obtains $D_a$ and an $\mathrm{EOO} = E_{sk_a}(h_a)$, where
$h_a = h(D_a)$. This evidence does not assure itself that $P_b$ is
the intended receiver. In other words, if the exchange would have
been carried out between parties $P_a$ and $P_c$, then the EOO
received by $P_c$ would have been identical (assuming that the same
symmetric key, $k_a$ is used). This way, once $P_b$ owns $D_a$ and
EOO, he might provide another party, $P_c$, with both items by using
a traditional channel. As a result, $P_c$ possesses the e-goods
coupled with a valid EOO for her. Party $P_a$, on the other hand,
does not own a receipt issued by $P_c$. Consequently, the protocol
neither provides fairness for $P_a$.

%
%.........................................................................
\subsection{On the security of a modified RSA-CEGD}\label{Sec:OtherVersion}
%.........................................................................
%
In \cite{NZCG05}, Nenadi\'c \emph{et al.} presented a different
version of the RSA-CEGD protocol with slight modifications. The
structure of this new proposal remains unaltered with respect to the
original version. In particular, items sent by $P_b$ during step E2
are the same that appears in the protocol here studied, i.e. the
VRES, the authorization token, and $P_b$'s certificate. Clearly, the
attacks described above are still applicable for this version.

%
%------------------------------------------------------------------------------
\section{Conclusions}\label{Sec:Conclusions}
%------------------------------------------------------------------------------
%
In this paper, we have demonstrated how the RSA-CEGD protocol
suffers from severe vulnerabilities. Our attacks show up that this
scheme can lead to an unfair situation for any of the two parties
involved in the exchange. To the best of our knowledge, the
aforementioned weaknesses have not been pointed out before.

%
%------------------------------------------------------------------------------
% Bibliography
%------------------------------------------------------------------------------

%------------------------------------------------------------------------------

\end{document}